# Some Experiences on BEPCII SRF System Operation

Huang Tong-ming[1;1)], Lin Hai-ying[1], Sha Peng[1], Sun Yi[1], Pan Wei-min[1], Wang Guang-wei[1], Dai Jian-ping[1], Li Zhong-quan[1], Ma Qiang[1], Wang Qun-yao[1], Zhao Guang-yuan[1], Mi Zheng-hui[1]

*1(Institute of High Energy Physics, CAS, Beijing, 100049, China)*

**Abstract** The Superconducting Radio Frequency (SRF) system of the upgrade project of Beijing Electron Positron Collider (BEPCII) has been in operation for almost 8 years. The SRF system has accelerated both electron and positron at the design beam current of 910 mA successfully, and a high beam intensity colliding of 860 mA (electron)*910 mA (positron) has been achieved in April 2014. Many problems were encountered during the operation, among which some were solved and some remain unsolved. This paper will describe some experiences on BEPCII SRF system operation, including the symptoms, causes and solutions.

## 1 Introduction

The upgrade project of Beijing Electron Positron Collider (BEPCII) has been put into operation since the end of 2006. The machine was constructed for both high energy physics (HEP) and synchrotron radiation (SR) researches. Considering the advantages of superconducting cavity, such as larger accelerating gradient, low RF power consumption and transmitted-out of HOMs, superconducting cavity (SCC) has been used in the BEPCII storage ring RF system [1]. The RF system includes two independent sub-systems. Each one is composed of a 500 MHz SCC, a 250 kW klystron and a low level system. In collision mode, a 1.5 MV accelerating voltage and 150 kW RF power can be provided for positron ring (BPR) and electron ring (BER) separately. In the SR mode, the beam circulates in the outer ring, and the system can provide 2.0 MV accelerating voltage and 100 kW RF power. The main parameters of the SRF system are listed in Table 1.

Table.1 Main parameters of BEPCII SRF system [2]

| Parameters \ Operation mode | Collision mode | SR mode |
|---|---|---|
| Frequency(MHz) | 499.8 | 499.8 |
| Cavity voltage(MV) | 2*1.5 | 2.0 |
| Energy loss per turn(keV) | 2*135 | 386 |
| Beam current(mA) | 2*910 | 250 |
| Beam power(kW) | 2*123 | 97 |
| Synchrotron phase (Degree) | 175 | 165 |
| Cavity number | 2*1 | 1 |
| Klystron number | 2*1 | 1 |
| Klystron power(kW) | 2*250 | 250 |
| Phase stability $\Delta\phi$ (Degree) | ±1.0 | ±1.0 |
| Amplitude stability $\Delta V_a/V_a$ (%) | ±1.0 | ±1.0 |

The SRF system has successfully accelerated both electron and positron at the design beam current of 910 mA, and a high beam intensity colliding of 860 mA (electron)*910 mA (positron) has been achieved in April 2014. However, many problems were encountered during the operation, e.g. overheating of the power coupler, frequent tripping at high beam intensity colliding, excessive helium pressure of the BER SCC and so on, in which some were solved and some remain to be resolved. This paper will describe some experiences on BEPCII SRF system operation, including the symptoms, causes and solutions.

huangtm@ihep.ac.n

## 2 Power coupler overheating

Each SCC is fed through one power coupler. As shown in Table.1, each coupler has to deliver 150 kW continuous wave (CW) RF power. The coupler is coaxial type, which consists of three parts: (1) a doorknob to realize the transition from waveguide to coaxial line; (2) a RF window to provide RF-transparent vacuum barrier; (3) a 50 Ohm coaxial line to transfer and feed RF power into cavity [3]. Overheating of the power coupler has always become a troublesome problem.

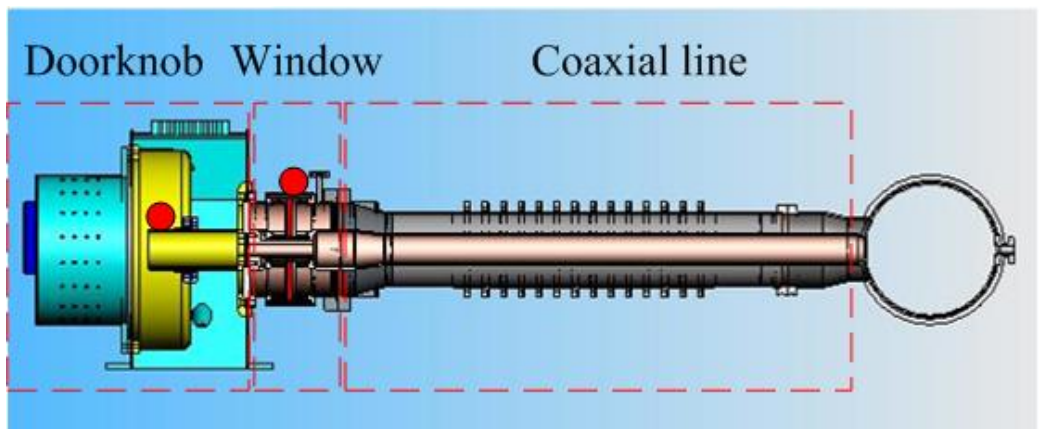

Fig.1. The 3D model of the power coupler: the area noted with red point is overheating.

During the annual operation from 2010 to 2011, the doorknob temperature noted with red point as shown in Fig.1 increased above 46 ℃ at the power level of 100 kW. Power leak was suspected at the beginning. So a reinstallation of the doorknob was implemented. However, the temperature was still too high. Finally, two big air conditioners were added for the doorknob air cooling. The temperature of the inlet cooling air reduced more than 10 ℃ and the doorknob temperature decreased accordingly [4]. Based on some experiments, the temperature of the doorknob cooling air should be below 20 ℃ for safety operation.

In Nov. 2013, the temperature of the window for the BPR power coupler was found excessively high, which became a big potential danger for the BPR SCC. In the worst case, the window temperature noted with red point as shown in Fig.1 increased to 50 ℃ in 5 minutes at the power level of 3 kW. Based on a series of experiments, the following characteristics were found: 1) the heating was very sensitive to the detune angle. As shown in Fig.2, at the power level of 40 kW, the temperature increased from 35 ℃ to 54 ℃ when the detune angle changed from 20 degree to -20 degree; 2) the heating is not uniform. The temperature of one side was 10 ℃ higher than the opposite side at the same power level of 80 kW. Since the window structure is axis-symmetric, we can deduce that the abnormal heating source is located around one side of the planar ceramic; 3) after three times 150 ℃ baking, the temperature increasing slope reduced obviously, as can be seen from Fig.3, which proved that the heating source isn't stable; 4) the heating usually became more serious if some big outgassing happened; 5) the vacuum of the overheating window has always been half order worse compared with the BER Ring normal window, which is more likely to occur discharging. So we speculated the possible overheating reason as follows: First, some serious discharging produced a large number of ions, then these ions bombarded the copper surface, and then the sputtered copper reacted with the residual gas of CO, $CO_2$, $H_2$ or $H_2O$, formed some kind of copper compounds of $CuC_2$ or CuH. Finally, the copper compounds deposited on the ceramic surface, which resulted in the ceramic overheating. If the above guess is reasonable, it tells us that keeping a good vacuum is very important for power coupler safety. Up to now, the problem is just alleviated instead of eliminated. A close monitoring of the window temperature has been done during current beam operation.

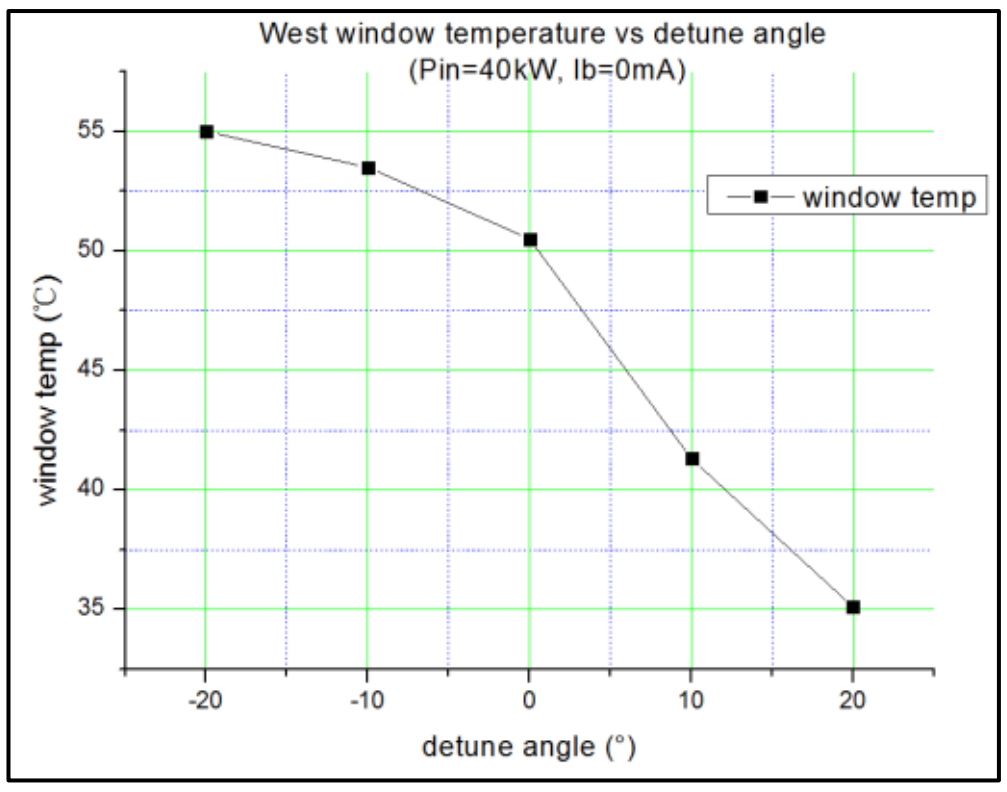

Fig.2. The window temperature is sensitive to the detune angle.

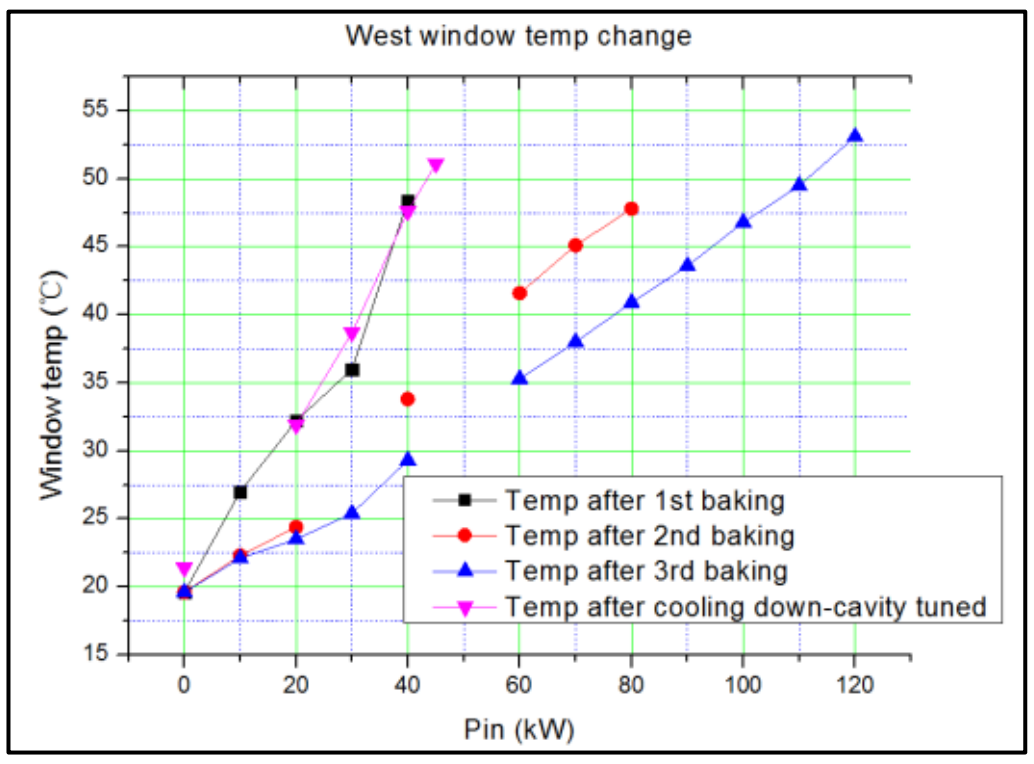

Fig.3. 150 ℃ baking is effective to alleviate the window overheating problem.

## 3 Frequent tripping during high beam intensity colliding

Frequent tripping during high beam intensity colliding, resulted in single beam or two-beam lost, has always been a stubborn problem. Sometimes the tripping happens at the beginning, sometimes in the medium and the latter of colliding. The tripping significantly limits the current intensity and the efficiency of colliding. A multi-channels oscilloscope is used to record the signals of beam, accelerating voltage, RF power, vacuum pressure, RF phase, tuner and LLRF control loop. With the trigger function, the signals changing within micro-seconds of tripping can be recorded, saved with data and shown with picture (Fig.4). Then, a preliminary tripping reason analysis can be done by distinguishing the time sequences and variation tendencies of the signals. A bunch by bunch beam diagnostic system has also been built to analyze the tripping reason by monitoring the variation of the beam itself [5]. The analysis results of the two methods can be compared and always agree with each other well. Up to now, almost all the tripping can be easily distinguished as beam instability, RF tripping, magnet and other system breakdown such as power supply and cryogenic, which is very helpful to reduce the tripping rate. Fig.5 shows the categorized pie chart of tripping rate during 2013 to 2014 annual operation based on the above tripping diagnostic methods.

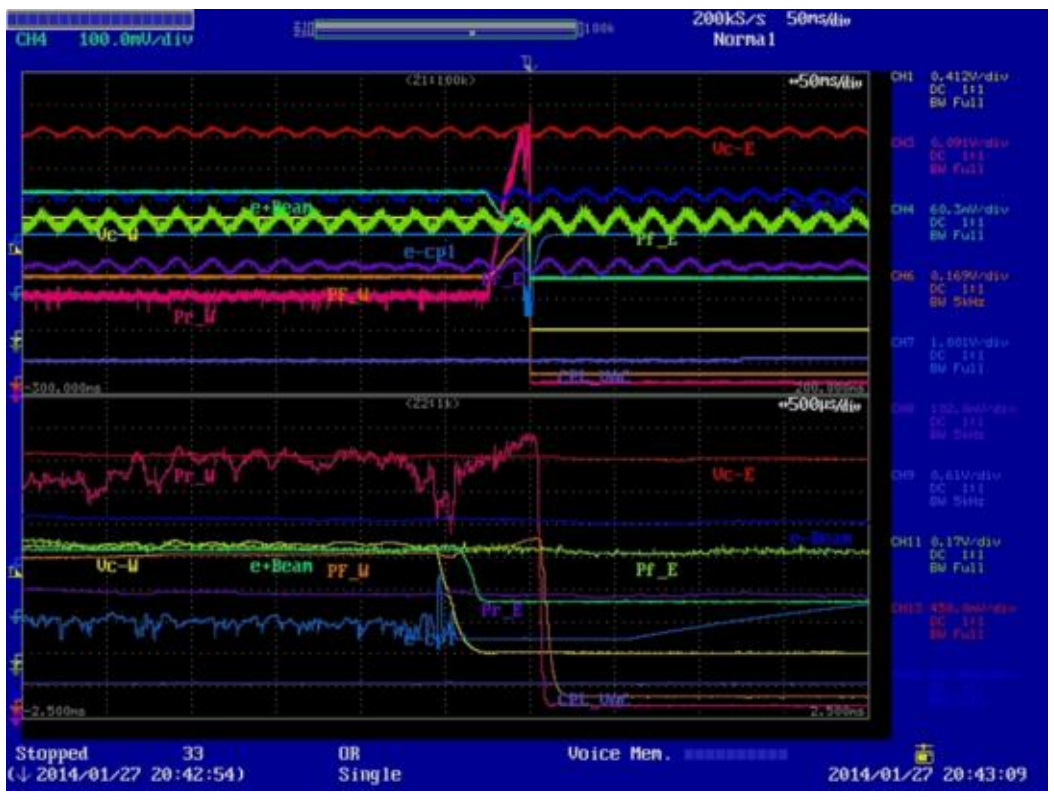

Fig.4. The signals variation during tripping, which recorded by the multi-channels oscilloscope.

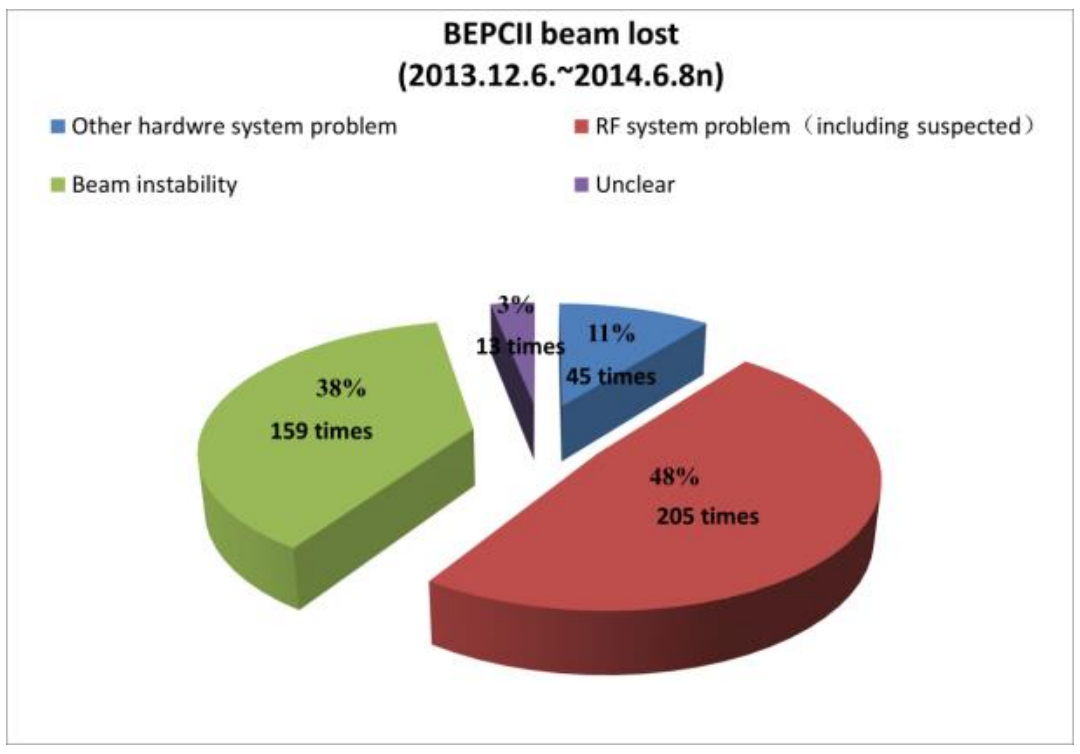

Fig.5. The categorized pie chart of tripping rate during 2013 to 2014 annual operation.

## 4 Other unsolved problems

Though most problems have been solved, there are still some unsolved stubborn problems in SRF system. Among them, false arc interlock trigger and excessive helium pressure are two typical ones.

**4.1 False arc interlock trigger**

Two arc detectors are used to protect the power couplers, one belongs to BPR ring, and the other one belongs to BER ring. Theoretically, the arc detector trigger only relates with the belonging ring. However, it sometimes triggered once the other ring tripped. What's more stranger is that there is no accompanied outgassing for more than 90% such kind of arc triggers. A series of methods have been done, such as improving the arc detector circuit to raising its anti-jamming capability and increasing the trigger threshold value. However, the problem still exists. An important information is that the arc detector never triggered if the tripping happened under no beam condition. So now the biggest suspect is that the fiber is disturbed by the beam lost induced X-ray. A shielding of the fiber will be tried soon.

**4.2 Excessive helium pressure of the BER SCC**

The excessive helium pressure of the BER SCC began in 2008, which has the following characteristics: 1) the helium pressure increasing only happens in the BER SCC and in collision mode, i.e. never happens in BPR SCC and SR mode; 2) the helium pressure increases quickly during the electron injecting, as shown in Fig.6; 3) the vertical orbit has an impact on the helium pressure; 4) the helium pressure just increases with the beam intensity, has nothing to do with the beam bunches; 5) the temperatures around the taper of both upstream and downstream increases accordingly; 6) the unknown heat source is about 80 W at the electron current of 750 mA [6]. By adjusting vertical orbit and increasing the helium pressure up-limit value, the problem has been alleviated to a certain extent now. However, it still exists; and more work from the view of the physics and engineering should be done to solve the problem.

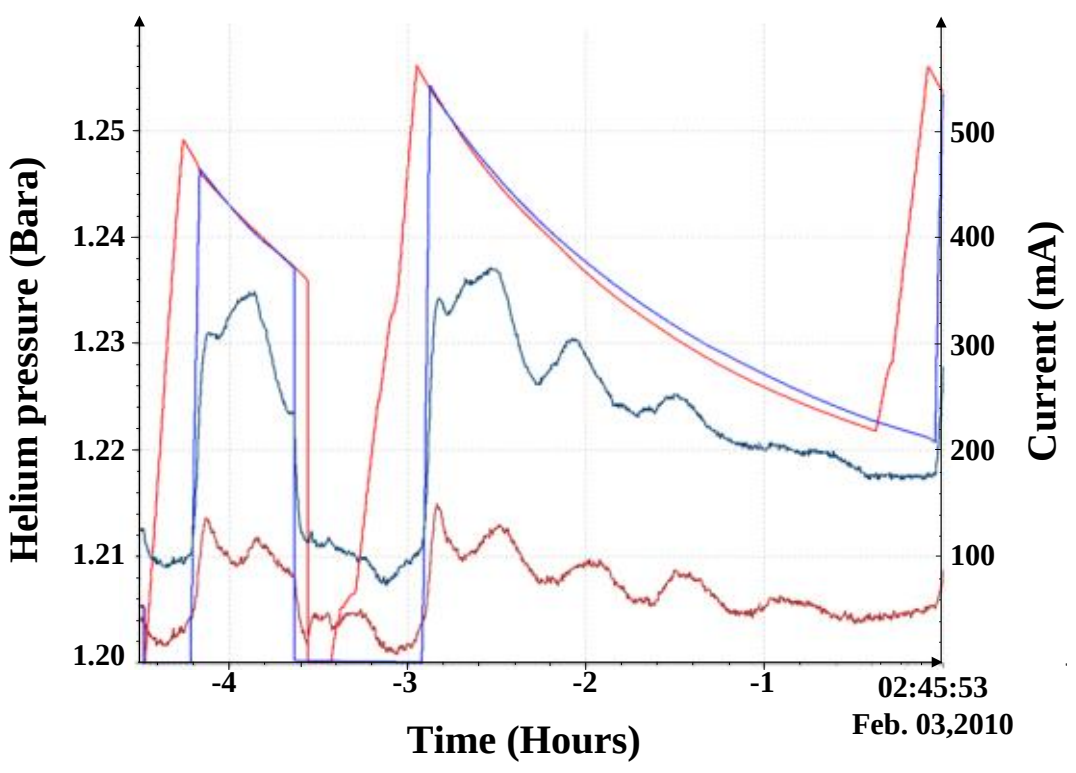

Fig.6. The helium pressure of the BER SCC increases quickly as the electron injecting (red: positron current; blue: electron current; indigo: BER SCC helium pressure; brown: BPR SCC helium pressure).

## 5 Summary

The SRF system for BEPCII has been in operation for almost 8 years and has successfully accelerated both electron and positron at the design beam current of 910 mA. During the operation, many problems were encountered. The doorknob of the power coupler was overheating and solved by reinforcing the air cooling; the window temperature of the BPR SCC power coupler is too high due to the damage of the ceramic caused by discharging, which tells us a good vacuum pressure is very important for coupler safety; Frequent tripping during high beam intensity colliding has always been a stubborn problem and a tripping diagnostic system has been built, which is very helpful to reduce the tripping rate; However, some problems are still open, such as false arc interlock trigger and excessive helium pressure of the BER SCC, which limits the operation efficiency and the beam intensity.

# BEPCII 超导高频系统运行中的一些经验

黄彤明[1; 1)]， 林海英[1]，孙毅[1]，潘卫民[1]，王光伟[1]，戴建枰[1]，李中泉[1]，马强[1]，王群要[1]，沙鹏[1]，赵光远[1]

1（中国科学院高能物理研究所 北京 100049）

**摘要：** BEPCII 超导高频系统已经运行将近 8 年。该系统总体运行情况良好。目前，正负电子束流的流强达到了 910 mA 的设计值；于 2014 年 4 月成功实现了正电子 910mA、负电子 860mA 的高流强对撞。然而，超导高频系统在实际运行中遇到了很多问题，也积累了很多经验。其中有些问题已经解决，有些问题尚未得到解决。本文将详细 BEPCII 超导高频系统在运行中碰到的一些典型问题，包括现象描述、原因分析和解决方法。